\documentclass[12pt,preprint]{aastex}

\def\dist{(m-M)_0}
\def\logz{\lbrack\hbox{M/H}\rbrack}
\def\feh{\lbrack\hbox{Fe/H}\rbrack}
\def\oh{\lbrack\hbox{O/H}\rbrack}
\def\ofe{\lbrack\hbox{O/Fe}\rbrack}
\newcommand{\mean}[1]{\langle #1 \rangle}

\slugcomment{To appear in the Astronomical Journal}
\shorttitle{Leo A RR Variables}
\shortauthors{Dolphin et al.}

\begin{document}

\title{Variable Stars in Leo A: RR Lyraes, Short-period Cepheids, and Implications on Stellar Content}

\author{Andrew E. Dolphin, A. Saha, and Jennifer Claver}
\affil{Kitt Peak National Observatory, National Optical Astronomy Observatories\linespread{1.0}\footnote{
NOAO is operated by the Association of Universities for Research in Astronomy, Inc. (AURA) under cooperative agreement with the National Science Foundation.  The WIYN Observatory is a joint facility of the University of Wisconsin-Madison, Indiana University, Yale University, and the National Optical Astronomy Observatories.}, P.O. Box 26372, Tucson, AZ 85726}
\email{dolphin@noao.edu, saha@noao.edu, jclaver@noao.edu}

\author{Evan D. Skillman}
\affil{Astronomy Department, University of Minnesota, Minneapolis, MN 55455}
\email{skillman@astro.umn.edu}

\author{A.A. Cole}
\affil{Physics \& Astronomy Dept., 538 LGRT, University of Massachusetts, Amherst, MA 01003}
\email{cole@condor.astro.umass.edu}

\author{J.S. Gallagher}
\affil{University of Wisconsin, Dept. of Astronomy, 475 N. Charter St, Madison, WI 53706}
\email{jsg@astro.wisc.edu}

\author{Eline Tolstoy}
\affil{Kapteyn Institute, University of Groningen, P.O. Box 800, 9700AV Groningen, the Netherlands}
\email{etolstoy@astro.rug.nl}

\and

\author{R.C. Dohm-Palmer and Mario Mateo}
\affil{University of Michigan, Dept. of Astronomy, 821 Dennison Building, Ann Arbor, MI 48109-1090}
\email{rdpalmer@astro.lsa.umich.edu, mateo@astro.lsa.umich.edu}

\begin{abstract}
We present the results of a search for short-period variable stars in Leo A.  We have found 92 candidate variables, including eight candidate RR Lyrae stars.  From the RR Lyraes, we measure a distance modulus of $\dist = 24.51 \pm 0.12$, or $0.80 \pm 0.04$ Mpc.  This discovery of RR Lyraes confirms, for the first time, the presence of an ancient ($> \sim 11$ Gyr) population in Leo A accounting for at least 0.1\% of the galaxy's $V$ luminosity.  We have also discovered a halo of old ($> \sim 2$ Gyr) stars surrounding Leo A, with a scale length roughly 50\% larger than that of the dominant young population.

We also report the discovery of a large population of Cepheids in Leo A.  The median absolute magnitude of our Cepheid sample is $M_V = -1.1$, fainter than 96\% of SMC and 99\% of LMC Cepheids.  Their periods are also unusual, with three Cepheids that are deduced to be pulsating in the fundamental mode having periods of under 1 day.  Upon examination, these characteristics of the Leo A Cepheid population appear to be a natural extension of the classical Cepheid period-luminosity relations to low metallicity, rather than being indicative of a large population of ``anomalous'' Cepheids.  We demonstrate that the periods and luminosities are consistent with the expected values of low-metallicity blue helium-burning stars (BHeBs), which populate the instability strip at lower luminosities than do higher-metallicity BHeBs.
\end{abstract}


\section{Introduction}
The Local Group dwarf irregular galaxy Leo A has been a source of considerable controversy, regarding both its distance and stellar content \citep{van00}.  Discovered by \citet{zwi42}, its first distance was determined by \citet{dem84} from photographic photometry.  They found an unusual CMD, apparently containing no normal supergiants (blue or red).  A large number of very blue ($B-V \sim -0.4$) stars were observed; these were interpreted as $60-120 M_{\odot}$ main sequence stars and thus evidence of very recent star formation.  No red supergiants were observed.  They used the three brightest blue stars to derive a distance of $\dist = 26.8 \pm 0.4$.

A later photographic study by \citet{san86} found significant discrepancies between the two photometry sets, strongly nonlinear (especially in $B$) in the sense that the photometry of \citet{dem84} was fainter than that of \citet{san86} for bright stars and brighter for faint stars.  \citet{san86} thus found a ``normal'' CMD, with both red and blue supergiants falling in locations similar to those of other dwarf irregular galaxies.  He therefore was able to use both the brightest three red star and the brightest three blue star methods to determine a distance modulus of $\dist = 26 \pm 1$.

The photometric distance determinations were confirmed by \citet{hoe94}, who determined a value of $\dist = 26.74 \pm 0.22$ based on four Cepheids.  \citet{tol96} then used their photometric data and distance to make the first stellar population analysis of Leo A.  Modeling the strong blue and red plumes in the CMD as blue and red supergiants, she concluded that Leo A has no very young stars and is apparently in a period of diminished star formation.

This picture of Leo A's distance and stellar content came into question with the HST-based study of \citet{tol98}.  With these data, it became clear that what had been interpreted by \citet{tol96} as a red supergiant sequence was actually a very narrow red giant branch, and that the broadening at its base was a vertically-extended red clump.  The distance modulus was therefore revised to $\dist = 24.2 \pm 0.2$, making Leo A's brightest supergiants and Cepheids two magnitudes fainter than had previously been believed.  The explanation for the error in the brightest supergiant distance is straightforward.  For dwarf galaxies, the luminosity of the brightest star is limited by the stellar IMF and small number statistics rather than stellar astrophysics \citep{roz94}.  Thus, while two giant galaxies can be reasonably expected to have similar brightest supergiants, this is not true of dwarfs, in which the same supergiants could be interpreted as a small nearby system or as a large distant system with equal validity.  This degeneracy was recognized by \citet{san86}, and was the reason for his 1 magnitude uncertainty.

However, the results of \citet{hoe94} have remained a puzzle.  Even with the shorter distance, there is no doubt that Leo A contains a large amount of recent star formation, and therefore should contain a significant Cepheid population.  As noted by \citet{tol98}, only one of the periodic variables found by \citet{hoe94} falls on the Cepheid period-luminosity (P-L) relation for the shorter distance.  The implication of these findings was therefore that Leo A is an unusual system -- one containing a significant amount of young stars and possibly an entirely-young system \citep{tol98}, but apparently without Cepheids.

By bringing the galaxy a factor of three closer, the blue supergiants previously assumed to be rather young were made older, thus producing the interpretation that most of Leo A's stars were formed $0.9 - 1.5$ Gyr ago, with a much smaller amount at more recent ages and perhaps none at older ages.  The conclusion of \citet{tol98} that any ancient population is very weak or perhaps non-existent was based on the relative weakness of the red giant branch, when compared with the red clump.  However, they noted that studies of Leo A main sequence turnoff stars or RR Lyrae stars would be needed in order to definitively establish the existence or lack of an ancient population.  With a distance of less than 1 Mpc, it is now feasible to search for RR Lyraes with ground-based telescopes.

The present work is a new study of variable stars in Leo A.  Our observations, photometry, and variable star search is described in Sections 2 and 3.  Our variable star results are discussed in Sections 4 and 5, while a stellar population gradient is examined in Section 6.  Section 7 contains the summary.

\section{Observations and Reduction}

Observations of Leo A were obtained at the WIYN 3.5m telescope on the nights of 20$-$22 December 2000, using the MIMO camera (described by \citet{sah00}).  Our observations are summarized in Table \ref{tab_obslog}.  The MIMO camera contains two 2k$\times$4k chips, separated by 10 arcsec.  Because of its size, shape, and orientation, Leo A fit completely on chip 2, allowing us to ``hide'' a bright star in the gap between the chips and obtain a halo field on chip 1.  A deep $V$ image is shown in Figure \ref{fig_image}.  Because of scheduling limitations, Leo A rose during the middle of the night, restricting our observing to a 6.5 hour window between its rise and morning twilight.
\placetable{tab_obslog}
\placefigure{fig_image}

As indicated in Table \ref{tab_obslog}, we had sub-arcsecond seeing for the entire run, with seeing as good as $0.5 - 0.6$ arcsec (excellent image quality is typical at WIYN).  This was was critical for our project, both in terms of reducing blending and reducing the effective background level.  As the RR Lyraes were near the limits of our photometry, they could not have been detected with significantly worse seeing.

To ensure a robust final result, the data were processed completely independently by two groups.  AED used his own package, an adaptation of HSTphot \citep{dol00a}, which includes image processing as well as photometry programs.  The reduction algorithms were similar to that implemented in IRAF, but include a more accurate overscan correction algorithm and an algorithm to mask pixels affected by crosstalk between MIMO's amplifiers.  Deep frames in $V$ and $R$ were created by aligning all images for the filter (23 $V$, 5 $R$) to the nearest pixel and coadding.  Photometry was determined for the deep frames, with the resulting star list used as a template (after determining appropriate coordinate transformations) for each of the individual epochs.  By restricting the star positions, we obtain more accurate photometry in the individual epochs, as each star has only one free parameter (brightness) instead of three (X, Y, and brightness).

JC and AS made a parallel reduction using IRAF and photometry using DoPHOT \citep{sch93}.  In this reduction, the images were adjusted prior to IRAF reduction, correcting for overscan and crosstalk using an IDL routine written by AS.  The normal IRAF procedure was then used to reduce the data.  The DoPHOT photometry consisted of two passes, similar in concept to that made by AED.  In the first pass, each individual-epoch image was photometered, along with the deep frames.  Coordinate transformations were made between the deep and individual star lists, and photometry was re-run on the individual frames.  A comparison between our deep photometry of Chip 2 is shown in Figure \ref{fig_compare}; there are no obvious inconsistencies between the two reductions.
\placefigure{fig_compare}

Because only one of the three nights was photometric, we could not make independent photometric solutions for each night.  Instead, we observed two standard fields -- SA 92 \citep{lan92} and NGC 2419 \citep{ste00} -- and used Leo A observations at similar airmass (n2086 and n2083; see Table \ref{tab_obslog}) to define secondary standards in our Leo A field.  Unfortunately, we did not observe SA 92 in $R$, and the \citet{ste00} $R$ standards in NGC 2419 only cover the region north of NGC 2419 (corresponding to chip 1 on our images).  Thus we were forced to use the chip 1 $R$ transformation for chip 2 also, under the assumption that the color term of the $R$-band transformation is the same in the two chips.  Data for other WIYN+MIMO studies has shown this to be nearly true; we increase our $R$ error bars by $0.02$ magnitudes to account for this source of uncertainty.  The resulting photometric calibration is accurate to $\pm 0.02$ magnitudes in $V$ and $\pm 0.04$ magnitudes in $R$.

Using our secondary standards within the Leo A images, we calibrated both photometry sets by forcing the mean difference between the calibrated and PSF-fit magnitudes of all secondary standards to be zero.  This resulted in a large photometric database consisting of the deep positions and magnitudes of all stars, as well as the magnitudes determined for each epoch.

\section{Variable Star Identification}

Variable stars were identified according to the procedure described by \citet{dol01b}, except that only the $V$ observations were used for locating variables.  Because of smaller error bars in the photometry, we adopted AED's PSF-fitting photometry as our primary measurement.  We believe that the smaller errors in AED's photometry resulted because his PSF formulation (an analytical function plus a residual image) permitted a more accurate modeling of the oddly-shaped PSFs (due to tracking errors) found in our data than does the DoPHOT PSF formaultion (analytical function only).  The measured magnitudes were nearly identical in the two reductions as noted above; however the smaller (and more accurate) error bars were necessary to accurately identify variable stars.

For a star to be classified as a variable star, it had to meet a strict set of criteria.  The first criterion was that the PSF-fitting photometry had to detect and classify it as a well-fit star in the deep $V$ frame.  We define ``well-fit'' to mean that the $\chi^2$ of the fit had to be 4.0 or less, the sharpness had to be between $-0.25$ and $0.25$ (a sharpness of zero is a perfect star), and the signal-to-noise ratio had to be 5.0 or more.  Additionally, the star had to be well-photometered in at least 12 of the 23 $V$ epochs and at least one $R$ epoch, and contamination from bright neighbors had to contribute no more than 20\% of the star's light within a PSF-sized aperture.

Stars that met these criteria had to pass additional variability tests.  First, to eliminate non-variable stars, we required that the reduced $\chi^2$ of the $V$ observations was 2.0 or greater and the standard deviation of the $V$ measurements was at least 0.14 magnitudes.  The first criterion eliminates stars that appear variable because their measurements have large uncertainties; the second eliminates stars where small systematic errors in the measurements could cause false detections.  Additionally, in case a star was spuriously flagged because of one or two bad points, we eliminated 1/3 of the points contributing the most to $\chi^2$ and required that this ``robust'' $\chi^2$ value was 0.5 or more.

Our final automatic variability test for periodicitiy and was made using the Lafler-Kinman \citep{laf65} $\Theta$ statistic as implemented by \citep{sah90}.  For a variety of trial periods, the value $\Theta(p)$ was calculated using
\begin{equation}
\Theta(p) = \frac {\sum_{i=1}^{N}(V_i - V_{i+1})^2}{\sum_{i=1}^{N}(V_i - \overline{V})^2},
\end{equation}
where $N$ is the number of epochs in which the star was observed, $V_i$ is the $V$ magnitude at epoch $i$, and $\overline{V}$ is the mean $V$ magnitude.  For uncorrelated data, $\Theta$ will be $\approx 2$, since the typical difference between any two data points would be $\sqrt{2}$ times the standard deviation (the denominator).  However, if the data are periodic and the correct trial period is used, $V_i$ and $V_{i+1}$ will be correlated, thus reducing $\Theta$.  We required $\Theta \le 1.0$.

Finally, we visually examined the image and light curves of each star, eliminating any candidate variables that either appeared non-stellar or non-periodic.

We found three additional variable stars from the DoPHOT photometry that were not recovered in our initial variable star search.  (C2-V09 had $\sigma V = 0.13$, C2-V30 had $\chi^2 = 4.4$ in the photometry, C2-V57 had $\sigma V = 0.12$.)  After visual inspection, these three stars were added to our list of variables.  In all, we found 92 variable stars, of which 82 have very good light curves.

Our list of variable stars is presented in Table \ref{tab_variables}, finding charts in Figure \ref{fig_charts}, and light curves in Figure \ref{fig_curves}.  The values given in Table \ref{tab_variables} are the variable star ID, chip and position (the FITS file from which the positions are taken will be published in the electronic AJ), period-averaged $V$ and $R$ magnitudes \citep{dol01b}, period, and light curve quality on a scale of 0 being the worst to 4 being the best \citep{dol01b}.  We note that our relatively short observing window created a large number of stars with possibly-aliased periods; these are indicated in Table \ref{tab_variables} and Figure \ref{fig_curves} by the same star ID being shown with multiple periods.
\placetable{tab_variables}
\placefigure{fig_charts}
\placefigure{fig_curves}

Only one of our variable stars (our C2-V25 = their V10) is in common with the list of \citet{hoe94}, albeit with a period of 1.4 days instead of 13 days.  Three of their other variables (V7, V9, and V13) are also clearly variable in our data.  However, none of the three passed our PSF fit cut of $\chi^2 \le 4.0$ (all were more extended than a typical star and thus are likely blends), so these three were excluded from our list.  Of their other objects, their V3 appears to be a galaxy in our images and V8 is also clearly elongated, thus being rejected as non-stellar objects by our photometry.  We see no sign of variability for the majority of their variables (V1, V2, V4, V5, V6, V11, V12, and V14); this is undoubtedly because of the sampling difference -- they had a baseline of a decade but only one pair of observations separated by less than 1 day, while we had a baseline of only 2.25 days but frequent observations during that period.  We note that the periods they determined for V2 (1.76 days), V11 (2.01 days), and V12 (0.51 days) are inconsistent with our data; we should have seen such variability during our 2.25 day baseline.  These period inconsistencies can be attributed to their poor sampling of time scales of a few days.  These concerns were raised in a later paper by the same authors \citep{sah96}, who noted the possibility that some of their Leo A Cepheids were long-period red variables.

A color-magnitude diagram showing our variable stars is given in Figure \ref{fig_cmd}, with the variable stars marked as circles (size representing the light curve quality).  We emphasize that colors and magnitudes were not used to determine our sample; the restriction of variable stars into two populations on the CMD is therefore real and not a selection effect.  We interpret the brighter population (shown as open circles) as classical Cepheids and the fainter population (shown as solid circles) as RR Lyraes.
\placefigure{fig_cmd}

\section{RR Lyraes}

Our primary goal in this project was a search for RR Lyraes.  Eight candidates were found.  An expanded CMD detailing this region is shown in Figure \ref{fig_cmd_rr}.  Two of these (C1-V03 and C1-V07) have extremely clean light curves, no period aliasing, were well-photometered and isolated from any source of potential contamination, and have colors that place them in the instability strip.  We therefore claim that these are \textit{bona fide} RR Lyraes.  C1-V03 has a period and light curve shape (rapid rise, slow descent) that make it appear to be an RRab.  C1-V07 has a shorter period and longer ascent time, making it appear to be an RRc.
\placefigure{fig_cmd_rr}

A third (C2-V79) is another good candidate.  Its red color appears to be the result of a bad point in $R$ as well as phase coverage; eliminating the point moves the star bluer by 0.3 magnitudes and accounting for the fact that it was observed in $R$ only near maximum moves it bluer by an additional 0.2 magnitudes.  We cannot make a type classification from its light curve, as it was not observed while ascending.  However, its long period would make it an RRab.  Finally, C2-V66 also appears to very likely be an RR Lyrae; its having two possible periods lowered it to a quality rating of 2 and makes a classification impossible.  As with C2-V79, it was only observed in $R$ near maximum and is thus bluer than its listed color.

We are less certain of the nature of the other four candidates.  C1-V01 has a period that seems too short to be an RR Lyrae (though it could be an RRd).  C2-V33 does not have the light curve shape one would expect to see at that period, perhaps indicating that it is an anomalous Cepheid rather than an RR Lyrae.  C2-V56 fits an RRc template light curve reasonably well at the period of 0.40 days, but a few points fall well off that template.  (Its red color is caused by poor phase coverage; it is likely $0.15-0.2$ magnitudes bluer than our measured color of $V-R = 0.42$.)  Finally, C2-V80 is our faintest candidate and thus has large error bars that prohibit a definitive classification.

Weighting by the light curve qualities of all eight candidate RR Lyraes, we measure a mean magnitude of $\mean{V} = 25.11 \pm 0.06$, where the quoted uncertainty is the uncertainty in the mean.  Using only the four best RR Lyraes, we find $\mean{V} = 25.08 \pm 0.05$.  Finally, the two \textit{bona fide} RR Lyraes produce a mean magnitude of $\mean{V} = 25.10 \pm 0.09$.  We adopt the final value, which is the most conservative of the values.  The DoPHOT magnitudes are not significantly different from these values.  Given that our completeness reached well beyond the RR Lyraes ($V \sim 25$) in the halo of Leo A (where all eight candidates are located), we find it unlikely that our calculated mean magnitude is biased.

In order to determine the distance, we adopt the RR Lyrae absolute magnitude calibration of \citet{car00}:
\begin{equation}
M_V = (0.18 \pm 0.09) (\feh + 1.5) + (0.57 \pm 0.07).
\end{equation}
We choose this calibration largely because its zero point is nearly identical to what we measured in IC 1613 ($M_V = 0.61 \pm 0.08$ at $\logz = -1.3 \pm 0.2$); see \citet{dol01b} for a more detailed discussion of the various RR Lyrae zero point values.  As is demonstrated below, the slope of the relation is irrelevant to this discussion.

Since no spectroscopy exists for stars in Leo A, the metallicity of Leo A RR Lyraes is very uncertain.  We can attempt to estimate this value with two methods.  First, we can fit the WFPC2 data of \citet{tol98} to the theoretical isochrones of \citet{gir00}, assuming an age of $10-15$ Gyr.  We find that the RGB color can be fit with a metallicity of $\logz = -1.8 \pm 0.3$.  The advantage of this determination is that it is based on old stars; the disadvantage is that metallicities measured by broadband photometry are less accurate than those determined spectroscopically.

An additional constraint is given by the present-day metallicity, as measured by HII region spectra ($\oh = -1.5 \pm 0.2$; van Zee et al. 1999).  Adopting an $\ofe$ ratio, we can use this as an upper limit to the metallicity of the RR Lyraes in Leo A.  While older stars in our galaxy (halo and globular cluster stars) at the metallicity of Leo A show $\alpha$ element enhancements of $\sim 0.4$ dex and those in the Draco, Ursa Minor, and Sextans dSph galaxies show enhancements of $\sim 0.3$ dex \citep{she01}, these enhancements are not seen in the young populations of other metal poor dwarf irregular galaxies, such as the SMC \citet{ven99} and NGC 6822 \citet{ven01}.  Thus we adopt an $\alpha$ element enhancement of $\ofe = 0.2 \pm 0.2$ dex, which results in a present-day metallicity of $\logz = -1.7 \pm 0.3$.

We believe the spectroscopic determination to be more accurate (given uncertainties in the isochrones), and thus adopt the value of $\feh = -1.7 \pm 0.3$, which results in an RR Lyrae absolute magnitude of $M_V = 0.53 \pm 0.08$.  Adopting an extinction of $A_V = 0.06$ \citep{sch98}, we measure a true distance modulus of $\dist = 24.51 \pm 0.12$, corresponding to a distance of $0.80 \pm 0.04$ Mpc.  Because the metallicity is very close to $\feh = -1.5$, the assumed slope in the RR Lyrae absolute magnitude calibration has little effect on the distance.  Adopting a steeper slope slope of 0.30 magnitudes per dex \citep{san93}, for example, would only increase the distance modulus by 0.02 magnitudes; adding this difference in quadrature to our $0.12$ magnitude uncertainty does not increase the uncertainty.

Our preferred distance is a factor of three smaller than that of $\dist = 26.74$ measured by \citet{hoe94}, but consistent with that of $\dist = 24.5 \pm 0.2$ determined from the RGB tip by \citet{tol98}.  (They were unsure as to whether or not a population older than 1.5 Gyr existed in Leo A, leading them to base their final distance of $\dist = 24.2 \pm 0.2$ largely on the blue loop and red clump positions than on that of the RGB tip, which is not a standard candle at ages younger than $\sim 2$ Gyr.)  We also note that their ``young RC'' distance of $\dist = 24.2 \pm 0.2$ is based on an assumed red clump absolute magnitude of $M_I \simeq -0.4$; the semi-empirical (and thus less model-dependent) technique described by \citet{dol01b} produces a more accurate absolute magnitude of $M_I = -0.67 \pm 0.1$ for the metallicity and age assumed by \citet{tol98}, thus increasing their red clump distance by $\sim 0.3$ magnitudes and making it consistent with the other distances.  This 15\% change in the Leo A distance could significantly alter the star formation history they determined; that problem should thus be revisited in light of our revised distance.

We note the possibility that the stars we are detecting are anomalous Cepheids rather than RR Lyraes.   We believe this to not be the case for several reasons.  As anomalous Cepheids tend to be $\sim 0.5$ magnitudes brighter than RR Lyraes, this would push the Leo A distance to $\dist = 25.0$ and would be inconsistent with the RGB tip and red clump distances quoted above.  Additionally, we do see short-period variables, discussed in the following section, which are about half a magnitude brighter than the RR Lyraes.

With the discovery of RR Lyrae stars in Leo A, we have definitively located a population of ``ancient'' ($> \sim 11$ Gyr; Walker 1989) stars in Leo A, a galaxy previously considered a potential candidate for a delayed onset of star formation \citep{tol98}.  A quantitative estimate of the extent of this population is extremely difficult to determine because of the completeness of our RR Lyrae survey, which is zero inside the region where young stars are present.  Scaling from the relatively uncrowded chip 1, we estimate a total of $\sim 50$ RR Lyraes in the entire galaxy, with a $1 \sigma$ lower limit of $\sim 10$.  Adopting M5's well-populated horizontal branch \citep{rei96} and its 65 known RR Lyraes \citep{kal00} as being typical of an ancient population, we estimate that such an M5-like population would account for $0.1\% - 1\%$ of Leo A's $V$ luminosity.  Instead using a globular cluster with proportionally fewer RR Lyraes (such as M13), the ancient population could be larger by an order of magnitude or more; we thus quote only a value of $0.1\%$ as the lower limit of the contribution.  Because of the large uncertainties in this calculation, an accurate quantification of the ancient population would require photometry reaching the ancient main sequence turnoffs ($M_V \sim +4$).

\section{Short Period Cepheids}

Although the primary goal of our project was the detection of RR Lyraes in Leo A, our 2-day baseline made us sensitive to any variable star with a period of 2 days or less.  It was perhaps surprising, given the lack of previously-known Cepheids in Leo A, that 91\% of our detected variable stars (84 out of 92) are brighter than $V = 24.5$, fall in or near the instability strip, and therefore apparently Cepheids.  Figure \ref{fig_cmd_ceph} shows our CMD, expanded around the area of the instability strip; small dots represent all stars, while the circles represent our variables.  The sizes of the circles indicate the light curve quality.  Note that the majority of objects lie near the intersection of the blue loop with the instability strip.  Variables with ambiguous periods, as well as those with quality ratings of 2 or lower \citep{dol01b}, are omitted, leaving 66 Cepheids with accurately-measured periods and excellent light curves.  We plot the P-L diagram of these variables in Figure \ref{fig_pl}.
\placefigure{fig_cmd_ceph}
\placefigure{fig_pl}

Because of the change in the P-L relation slope at a period of 2 days \citep{bau99}, we had to calculate P-L relations for short-period Cepheids like those seen in Leo A instead of using the published relations of \citet{mad91} or \citet{uda99a}.  To do this, we selected objects with periods of 2 days or less from the OGLE SMC Cepheid database \citep{uda99b} and fit the following relations for fundamental-mode and first-overtone pulsators:
\begin{equation}
M_V (\hbox{FM}) = -3.14 \log(P) - 1.04
\end{equation}
and
\begin{equation}
M_V (\hbox{FO}) = -3.39 \log(P) - 1.73.
\end{equation}
Using the distance modulus of $\dist = 24.51 \pm 0.12$ determined in Paper I from the RR Lyraes, we plot the P-L relations as solid lines on Figure \ref{fig_pl}.

Since we have insufficient phase coverage to attempt Fourier decomposition, we must classify Cepheids based simply on their positions in the P-L diagram.  From Figure \ref{fig_pl}, we classify 19 of our 66 Cepheids as fundamental-mode, 38 as first-overtone, and 2 as second-overtone.  The remaining 7 Cepheids fall between the relations, and thus cannot be definitively classified.

Of particular interest are three fundamental-mode Cepheids with periods of less than 1 day: C1-V08, C2-V49, and C2-V77.  With periods of $\sim 0.8$ days, these objects fall in the period regime generally populated by RR Lyraes and overtone-pulsating Cepheids.  That these objects are not RR Lyraes is clear from the fact that they are $\sim 1.4$ magnitudes brighter than the RR Lyraes discovered in Paper I.  The likelihood of their being overtone-pulsating Cepheids is also small, given that they fall 0.6 magnitudes below the overtone P-L relation in Figure \ref{fig_pl} but only 0.06 magnitudes away from an extrapolation of the fundamental-mode relation.  Thus we conclude that these objects are fundamental-mode Cepheids, despite having periods similar to those of RR Lyraes.

There is some ambiguity as to whether these objects should be classified as classical or anomalous Cepheids, as the anomalous Cepheid P-L relations of \citet{nem94} are very similar to those given above.  Adjusting their M15 distance to $\dist = 15.23$ \citep{rei99}, their relations become:
\begin{equation}
M_V (fundamental) = -3.13 \log(P) - 0.99
\end{equation}
and
\begin{equation}
M_V (overtone) = -3.13 \log(P) - 1.54.
\end{equation}
The fundamental-mode relations are clearly the same; the overtone relations are also consistent in the period range of \citet{nem94} once the slope differences are accounted for.  To further confuse matters, there is an overlap in period between the SMC classical Cepheids of \citet{uda99b} and the anomalous Cepheids of \citet{nem94}.  Thus no clear distinction between these two classes of objects appears to exist in terms of position in a P-L relation.  For the sake of clarity, we will define ``classical'' Cepheids as population I (young) stars, and ``anomalous'' Cepheids as population II (old) stars.

\subsection{The Period-Luminosity Relations}

A brief review of the origin of the P-L relations may help explain this apparent contradiction of finding Cepheids with periods normally associated with RR Lyraes.  What follows is only a back-of-the-envelope calculation, but is sufficiently accurate to demonstrate the nature of these objects.  The pulsation period ($P$) is related to the mean density ($\rho$) as follows:
\begin{equation}
\label{eq_per1}
P \propto \rho^{-1/2}.
\end{equation}
We can use the definition of density ($\rho \propto \frac{M}{R^3}$) and effective temperature ($T^4 \propto L / R^2$) to rewrite equation \ref{eq_per1} as
\begin{equation}
\label{eq_per2}
P \propto \frac{L^{3/4}}{M^{1/2} T_{eff}^3}.
\end{equation}
Since effective temperature as a function of luminosity is constrained to a small range within the instability strip, the pulsation period is essentially a function of luminosity and mass.

For evolving population I stars, evolutionary motion is approximately horizontal in the H-R diagram.  Therefore the main sequence mass-luminosity relation ($L/L_{\odot} \approx (M/M_{\odot})^{3.5}$) is roughly true also for pulsating variables, reducing equation \ref{eq_per2} to
\begin{equation}
\label{eq_popI}
P = C \frac{(L/L_{\odot})^{0.61}}{T_{eff}^3},
\end{equation}
where $C$ is a constant.  Incorporating the dependence of $T_{eff}$ on $L$ within the instability strip, one obtains the well-known Cepheid P-L relation.

This relation does not hold true for population II stars, however, which become significantly brighter after leaving the main sequence and arrive at the horizontal branch ($M_V \approx +0.6$, or $L \approx 50 L_{\odot}$) with a mass of only $\approx 0.6 M_{\odot}$.  With the mass thus fixed, the period should be given by:
\begin{equation}
\label{eq_popII}
P = 1.29 C \frac{(L/L_{\odot})^{0.75}}{T_{eff}^3},
\end{equation}
where $C$ is the same constant as in equation \ref{eq_popI}.

We note that this population II P-L relation has a very similar slope to the population I P-L relation given in \ref{eq_popI}, a feature observed by \citet{baa63} in their observations of M31.  Using equations \ref{eq_popI} and \ref{eq_popII}, we predict that a pop II variable with $L = 10^4 L_{\odot}$ (a typical luminosity of Baade's pop II variables) will have a period $\sim 5$ times that of a pop I variable of the same luminosity.  This corresponds to a difference of $\Delta \log P$ of $\sim 0.7$, consistent with the value of 0.8 measured by \citet{baa63}.

More pertinent to our discussion of Leo A, one can use these equations to determine the expected luminosity of a classical Cepheid with the same period of an RR Lyrae.  For a population I and a population II variable with the same period and effective temperature, we find
\begin{equation}
(L_{Ceph}/L_{\odot})^{0.61} = 1.29 (L_{RR}/L_{\odot})^{0.75} = 24.
\end{equation}
Using the mass and luminosity given above for the RR Lyrae and the main sequence mass-luminosity relation for the Cepheid, we find $L_{Ceph} \approx 190 L_{\odot}$, meaning that an RR Lyrae will be $\sim 1.45$ magnitudes fainter than a Cepheid of equal period and luminosity.  Even though the above calculations are rough approximations, it is interesting to note that this is almost exactly the magnitude difference observed in our data between the three variables in question ($\mean{V_{Ceph}} = 23.74$), which we have inferred to be fundamental-mode pulsators, and our RR Lyraes ($\mean{V_{RR}} = 25.11$).

\subsection{Metallicity Effects on the Cepheid Population}

While we have demonstrated that fundamental-mode Cepheids \textit{can} be found with the period and absolute magnitude claimed, it is somewhat disturbing that a population of such objects has not previously been observed.  The Cepheid P-L relations of \citet{uda99a} show LMC Cepheids reaching $M_V \approx -2.6$ in large numbers, and SMC Cepheids reaching $M_V \approx -1.4$.  Likewise, IC 1613 (a dwarf slightly more metal-poor than the SMC) shows Cepheids reaching $M_V = -1.3$ \citep{dol01a}.  First-overtone pulsators fall off at about the same absolute magnitudes in each of those data sets; however we observe significant numbers of both classes of Leo A Cepheids down to $M_V = -0.8$.  In fact, the median $V$ magnitude of the Cepheids in our sample is $M_V = -1.1$, which is fainter than 95.5\% of the OGLE SMC Cepheids and 98.9\% of their LMC Cepheids.  Note that this is not a completeness problem with the OGLE observations.  Their survey of the SMC shows a completeness (estimated from artificial star counts) of roughly 96\% at $I = 19.5$ ($M_I \approx +0.6$) for their lower-density fields and above 80\% for their densest field \citep{uda98}.  Comparison of stars in overlapping regions of their survey shows an overall 94\% completeness in their Cepheid catalog \citep{uda99b}.  Due to its smaller distance, the OGLE LMC catalog should be even less prone to completeness problems.  This is conceivably a selection effect in our Leo A data; however, if a significant number of longer-period Cepheids were present, they would have been found by \citet{hoe94}.

In addition to the low luminosities, another surprise was the large number of Cepheids recovered in Leo A.  With a luminosity 1.5\% that of the SMC, one would expect the 2083 OGLE Cepheids to scale to $\sim 30$.  Instead, we find three times that number.

The solutions to both of these dilemmas rest in metallicity dependency of the blue loop position on the CMD.  Synthetic CMDs for a variety of metallicities, calculated using the \citet{gir00} isochrones using constant star formation rates from 0 to 15 Gyr ago, are shown in Figure \ref{fig_synth}.  At solar metallicity, blue helium-burning stars (BHeBs) remain to the blue of the instability strip at magnitudes of $M_V < -4$.  This can be seen in the top left panel of Figure \ref{fig_synth}, where no stars fall in the instability strip.  At lower metallicity, however, BHeBs become increasingly blue.
\placefigure{fig_synth}

The effect of this metallicity dependence is twofold.  First, at lower metallicities, the intersection of the blue loop with the instability strip falls at fainter magnitudes.  This is apparent in Figure \ref{fig_synth}; the absolute magnitude is $M_V = -1.8$ at Z=0.004, $M_V = -0.9$ at Z=0.001, and $M_V = -0.4$ at Z=0.0004.  Secondly, because the blue loops are more strongly populated at fainter absolute magnitudes (a result of evolutionary timescales and the stellar IMF), more stars will fall within the instability strip for a low-metallicity system than for a high-metallicity system with equal star formation rates.

The presence of short-period Cepheids in young, low-metallicity populations is not a new discovery.  For example, the theoretical models of \citet{bon97} show this effect.  However, we note that such objects are ordinary post-main sequence population I stars and thus are properly classified as classical Cepheids and not as anomalous Cepheids.

\subsection{Star Formation History from the Cepheid Population}

A nagging question in our new variable star search is to explain the unsuccessful search of \citet{hoe94}.  Despite an 11-year baseline and observations spaced as close as 1 day apart, they failed to conclusively identify any Cepheids.  In contrast, we report the discovery of 84 Cepheid candidates based on three nights of data.

It is clear that \citet{hoe94} would have not been sensitive to the many $\sim 1$ day Cepheids that we find in abundance due to the BHeBs.  However, the population of the instability strip is nonzero at magnitudes brighter than $M_V = -2.3$.  Figure \ref{fig_synthtop} shows a more strongly-populated synthetic CMD, with a constant star formation rate from 0 to 1 Gyr ago and a metallicity of Z=0.0004; stars are present in the instability strip up to the top of the CMD.
\placefigure{fig_synthtop}

While it is possible that the lack of bright Cepheids may result from a recent decrease in the star formation rate, we do not find this to be essential.  Of the 5405 stars inside the instability strip of Figure \ref{fig_synthtop}, only 75 have absolute magnitudes brighter than $M_V = -2.5$ (corresponding to $P > 3 d$).  Scaling to the number of observed Cepheids in Leo A (84), we would expect only $\sim 1.2$ bright (long-period) Cepheids in Leo A because of both the stellar IMF and evolutionary timescales.  This is also seen in our CMD, which shows no clear evidence of stars brighter than $V = 22$ in the instability strip, although foreground contamination prohibits a definitive measurement.  Thus we find no evidence of an unusual star formation rate.  The recent ($<100$ Myr) star formation rate can not have been significantly higher than at older ages ($\sim 650$ Myr), in which case we would expect to see significant numbers of bright Cepheids, but Cepheid statistics provide no lower limit on the recent star formation rate.

This is not the first time that the absence of long-period Cepheids in a dwarf irregular galaxy has been pondered.  In their study of WLM, \citet{san85} found numerous Cepheids with periods between 3 and 10 days, but none with periods longer than 10 days.  They speculated that the absence of the longer period Cepheids could be due to either stochastic variations in the rate of star formation or (their favored explanation) a metallicity dependency of the lengths of the blue loop excursions of the post main sequence evolutionary tracks.  Later, \citet{ski89} measured HII region abundances in WLM, showing it to be more metal rich than Sextans A, which was known to have several longer-period Cepheids. Today we know that Sextans A has experienced a relatively high rate of star formation in the last 50 Myr \citep{doh97}, corresponding to main sequence turnoff masses in excess of 7 M$_{\odot}$ and thus Cepheids with periods in excess of 10 days, while the star formation rate in WLM has been relatively low during this period \citep{dol00b}.  Thus, given the relatively small numbers of total stars in the low-mass, low-metallicity dwarf irregular galaxies, small numbers of long-period Cepheid variables are expected.  Only enhanced recent star formation rates will produce statistically meaningful samples of long-period Cepheid variable stars in a single dwarf galaxy.

\section{Stellar Populations}

Although the primary topic of this paper is the variable star content of Leo A, we have found an interesting difference in the distributions of young and old stars in Leo A, with red stars ($22.0 < V < 23.5$ and $0.45 < V-R < 0.8$) distributed over a larger scale than blue stars ($21.5 < V < 23.5$ and $-0.3 < V-R < 0.2$).  The distributions of these populations are shown in Figure \ref{fig_contour}.  The population difference is also supported by the observation that one of out two \textit{bona fide} RR Lyraes (C1-V03) falls beyond the outermost contour of the blue stars and the other between the outermost and second contours.
\placefigure{fig_contour}

In order to better-quantify the population differences, we have divided the galaxy into three regions based on the blue star density.  The three resulting CMDs are shown in Figure \ref{fig_cmdsplit}.  The inner region shows a blue/red ratio of 0.87, the middle region a ratio of 0.25, and the outer region a ratio of 0.03.  We interpret this gradient as the sign of an old stellar halo surrounding the body of the galaxy, similar to those found in found in many other galaxies, such as WLM \citep{min97}, UGC 4483 \citep{dol01a}, and LGS 3 \citep{mil01}.  Comparing contour spacing, this stellar halo appears more compact than the HI halo found by \citet{you96}.  As our data were optimized for a variable star search rather than CMD analysis (thus giving us extremely deep photometry in $V$ and very shallow photometry in $R$), we cannot make a more detailed analysis of the populations.  We note that the population gradient we observe would require that such a study sample the entire galaxy in order to obtain a complete picture of the stellar populations and their spatial distributions.
\placefigure{fig_cmdsplit}

We note that our CMD shows a lack of bright supergiants in Leo A.  The brightest blue star falls at $V = 19.23$, or $M_V = -5.33$, corresponding to an age of 40 Myr according to the theoretical models of \citet{fag94} at Z=0.0004.  While the presence of HII regions demonstrates that Leo A contains a non-zero number of very young ($\le \sim 6$ Myr) main sequence stars, a large and steady amount of star formation should have produced blue supergiants roughly 2.5 magnitudes brighter than what is observed.  The brightest three blue stars have an average magnitude of $\mean{V} = 19.39$, or $M_V = -5.17$.  This is undoubtedly the cause of the incorrect distances measured by \citet{dem84} and \citet{san86}, as the brightest supergiant studies assumed absolute magnitudes of $M_V \sim -7$ for the brightest blue stars in Leo A.

\section{Summary}

We have presented the results of a search for short-period variables in Leo A, based on three consecutive nights of WIYN MIMO observations.  We have located 92 variable star candidates.  When placed on a CMD, the variable stars are clearly divided into two populations.  The majority of our variables are short-period classical Cepheids; eight are candidate RR Lyraes.  From the RR Lyraes, we measure a distance modulus of $\dist = 24.51 \pm 0.12$.  The discovery of RR Lyraes in Leo A disproves the hypothesis that the onset of star formation in Leo A was delayed until a few Gyr ago.

We have examined the Cepheid population of Leo A, which appears to be significantly different from previously-studied Cepheid populations.  We found 84 Cepheid candidates, of which 66 have excellent light curves in $V$ and $R$ and unambiguous periods.  Rather than a typical Cepheid population, with periods of 2-60 days and absolute magnitudes of $-2 < M_V < -6.5$, we found very faint ($-0.7 < M_V < -2.3$) Cepheids with short periods, including three fundamental pulsators with periods of under 1 day.  We have examined this unusual population in terms of pulsation theory, and have demonstrated that these objects fall at the expected positions on fundamental and overtone P-L relations.  The unusual Cepheid magnitudes are caused by the metallicity dependence of the blue loops.  We find that this also explains the relatively large number of Cepheids observed in Leo A (three times the number expected from scaling the SMC Cepheid population to Leo A's luminosity).  We believe that most of these unusual aspects of this Cepheid population should, in fact, be a common feature of low-metallicity star-forming galaxies.

We have also examined the distributions of blue stars (primarily blue He burning stars) and red stars (primarily red giants) in our image, and discovered an old stellar halo in Leo A that extends well beyond the extent of the young population.  A CMD of the halo looks typical for an old ($> \sim 2$ Gyr) population, with the only obvious feature being the red giant branch; a definitive age-dating of this population will only be possible with deeper photometry.

\acknowledgments
We would like to thank the NOAO TAC for the allocation of telescope time and the KPNO staff for their assistance.  EDS gratefully acknowledges partial support from NASA LTSARP grant NAG5-9221 and the University of Minnesota.

\clearpage
\begin{figure}
\caption{Combined $V$ image of Leo A.  North is to the left and East is down.  Each chip contains $2048 \times 4096$ pixels, with a scale of 0.141 arcsec per pixel.  The field is 9.6 arcmin on a side.}
\label{fig_image}
\end{figure}

\begin{figure}
\caption{Comparison between our two sets of chip 2 deep photometry.  We find no systematic differences of note.}
\label{fig_compare}
\end{figure}

\begin{figure}
\caption{Variable star finding charts.}
\label{fig_charts}
\end{figure}

\begin{figure}
\caption{Light curves of Leo A variable stars.}
\label{fig_curves}
\end{figure}

\begin{figure}
\caption{Color-Magnitude diagram of Leo A.  Variable stars are plotted as circles; circle size indicates the quality of the star's light curve (larger circles indicate higher quality).  Open circles indicate our brighter population of variables (Cepheids); filled circles our fainter population (RR Lyraes).  Error bars are taken from our photometry of the deep $V$ and $R$ frames.  Uncertainties in $R$-band magnitudes dominate the color error bars.}
\label{fig_cmd}
\end{figure}

\clearpage
\begin{figure}
\caption{Leo A CMD, expanded to show the RR Lyraes in detail.}
\label{fig_cmd_rr}
\end{figure}

\begin{figure}
\caption{Leo A CMD, expanded to show the Cepheids in detail.  Our mean $R$ magnitudes of the variables have more scatter than our $V$ magnitudes because of the limited phase coverage (5 epochs).}
\label{fig_cmd_ceph}
\end{figure}

\begin{figure}
\caption{Period-luminosity diagram of bright variables in Leo A.  Only Cepheids with unambiguous periods and quality ratings of 3 or 4 are shown.  The three short-period fundamental-mode Cepheids are marked as circles.  The lines are the P-L relations for short-period SMC Cepheids in the OGLE database \citep{uda99b}, assuming an SMC distance modulus of $\dist = 18.88$ \citep{dol01c}.  The lines are truncated near $\mean{V} = 23.4$ because the SMC does not contain a sufficient number of fainter Cepheids to constrain the P-L relations.  Note that the lack of variables with periods of a day is caused by our inability to measure unambiguous periods for such objects.}
\label{fig_pl}
\end{figure}

\begin{figure}
\caption{Synthetic CMDs based on the \citet{gir00} isochrones, calculated for the stated metallicities assuming a constant star formation rate.  The lines indicate the rough location of the instability strip.}
\label{fig_synth}
\end{figure}

\begin{figure}
\caption{Synthetic CMD based on the \citet{gir00} isochrones, calculated for Z=0.0004 ($\logz = -1.7$) with a constant star formation rate from 0 to 1 Gyr.  The lines indicate the rough location of the instability strip.}
\label{fig_synthtop}
\end{figure}

\begin{figure}
\caption{Density maps of (a) blue plume stars ($21.5 < V < 23.5$ and $-0.3 < (V-R) < 0.2$) and (b) red giants ($22.0 < V < 23.5$ and $0.45 < (V-R) < 0.8$), in arcsec$^{-2}$  Contour levels are plotted every 0.33 dex.  Note that the red star distribution is larger than the blue star distribution (as evidenced by the spacing between contours), indicating an old halo.}
\label{fig_contour}
\end{figure}

\begin{figure}
\caption{Color-Magnitude diagram of Leo A, divided by density of blue plume stars ($21.5 < V < 23.5$ and $-0.3 < (V-R) < 0.2$) as described in the text.}
\label{fig_cmdsplit}
\end{figure}

\clearpage
\begin{deluxetable}{lccccc}
\tablecaption{Observation Log. \label{tab_obslog}}
\tablewidth{0pt}
\tablehead{
\colhead{Image ID} &
\colhead{Epoch\tablenotemark{a}} &
\colhead{Exposure Time (s)} &
\colhead{Filter} &
\colhead{Airmass} &
\colhead{Seeing\tablenotemark{b}}}
\startdata
n1080 & 899.7859 & 1800 & $V$ & 1.90 & 0.86 \\
n1083 & 899.8156 & 1800 & $V$ & 1.54 & 0.77 \\
n1084 & 899.8390 & 1800 & $V$ & 1.36 & 0.81 \\
n1085 & 899.8623 & 1800 & $V$ & 1.23 & 0.68 \\
n1086 & 899.8861 & 1800 & $V$ & 1.14 & 0.58 \\
n1089 & 899.9353 & 1800 & $V$ & 1.03 & 0.66 \\
n1095 & 900.0157 & 1800 & $V$ & 1.02 & 0.65 \\
n1096 & 900.0390 & 1800 & $V$ & 1.05 & 0.81 \\
n2081 & 900.7897 & 1800 & $V$ & 1.81 & 1.01 \\
n2082 & 900.8130 & 1800 & $V$ & 1.54 & 0.88 \\
n2083 & 900.8367 & 1200 & $R$ & 1.36 & 0.79 \\
n2085 & 900.8577 & 1200 & $R$ & 1.24 & 0.72 \\
n2086 & 900.8742 & 1800 & $V$ & 1.17 & 0.72 \\
n2087 & 900.8975 & 1800 & $V$ & 1.10 & 0.71 \\
n2088 & 900.9210 & 1200 & $R$ & 1.05 & 0.72 \\
n2089 & 900.9374 & 1200 & $V$ & 1.02 & 0.79 \\
n2098 & 901.0233 & 1200 & $V$ & 1.03 & 0.75 \\
n2099 & 901.0397 & 1500 & $V$ & 1.06 & 0.75 \\
n3060 & 901.7571 & 1800 & $V$ & 2.36 & 0.96 \\
n3061 & 901.7804 & 1800 & $V$ & 1.90 & 0.76 \\
n3063 & 901.8102 & 1200 & $R$ & 1.54 & 0.66 \\
n3064 & 901.8266 & 1200 & $R$ & 1.41 & 0.58 \\
n3065 & 901.8429 & 1800 & $V$ & 1.30 & 0.54 \\
n3066 & 901.8662 & 1800 & $V$ & 1.19 & 0.51 \\
n3068 & 901.9130 & 1800 & $V$ & 1.05 & 0.57 \\
n3069 & 901.9364 & 1200 & $V$ & 1.02 & 0.61 \\
n3074 & 902.0052 & 1800 & $V$ & 1.02 & 0.53 \\
n3075 & 902.0286 & 1800 & $V$ & 1.05 & 0.55 \\
\enddata
\tablenotetext{a}{HJD$-$2451000.}
\tablenotetext{b}{Seeing values taken from our PSF-fitting photometry.}
\end{deluxetable}

\clearpage
\begin{table}
\begin{center}
\addtocounter{table}{1}
Table \thetable.  Photometry of Variable Stars. \label{tab_phot}
\begin{tabular}{cccccc}
\\
\tableline
\tableline
         &           C1-V01 &           C1-V02 &           C1-V03 &           C1-V04 &           C1-V05 \\
\tableline
HJD$^a$  &      $V$         &      $V$         &      $V$         &      $V$         &      $V$         \\
899.7859 & $25.232\pm0.241$ & $23.412\pm0.046$ & $25.307\pm0.257$ & $24.513\pm0.124$ & $23.653\pm0.057$ \\
899.8156 &      . . .       & $23.395\pm0.037$ & $25.549\pm0.264$ & $24.423\pm0.094$ & $23.659\pm0.048$ \\
899.8390 & $25.761\pm0.344$ & $23.365\pm0.038$ &      . . .       & $23.969\pm0.066$ & $23.611\pm0.048$ \\
899.8623 & $25.264\pm0.162$ & $23.437\pm0.031$ & $24.607\pm0.089$ & $23.705\pm0.040$ & $23.762\pm0.042$ \\
899.8861 & $24.916\pm0.094$ & $23.441\pm0.025$ & $24.640\pm0.073$ & $23.548\pm0.028$ & $23.706\pm0.032$ \\
\end{tabular}
\end{center}
$^a$ HJD$-$2451000
\end{table}
\clearpage
\begin{table}
\begin{center}
Table \thetable---Continued \\
\begin{tabular}{cccccc}
\\
\tableline
\tableline
         &           C1-V06 &           C1-V07 &           C1-V08 &           C1-V09 &           C1-V10 \\
\tableline
HJD$^a$  &      $V$         &      $V$         &      $V$         &      $V$         &      $V$         \\
899.7859 & $23.864\pm0.069$ & $25.229\pm0.241$ &      . . .       & $23.788\pm0.065$ & $23.094\pm0.034$ \\
899.8156 & $23.259\pm0.055$ & $25.219\pm0.193$ & $24.122\pm0.072$ & $23.989\pm0.064$ & $23.221\pm0.032$ \\
899.8390 & $23.683\pm0.052$ & $25.444\pm0.252$ & $23.951\pm0.066$ & $24.053\pm0.072$ & $23.288\pm0.036$ \\
899.8623 & $23.436\pm0.031$ & $25.707\pm0.243$ & $24.069\pm0.055$ & $24.062\pm0.055$ & $23.289\pm0.028$ \\
899.8861 & $23.246\pm0.021$ & $25.886\pm0.227$ & $23.950\pm0.039$ & $23.791\pm0.035$ & $23.321\pm0.023$ \\
\end{tabular}
\end{center}
\end{table}
\clearpage
\begin{table}
\begin{center}
Table \thetable---Continued \\
\begin{tabular}{cccccc}
\\
\tableline
\tableline
         &           C2-V01 &           C2-V02 &           C2-V03 &           C2-V04 &           C2-V05 \\
\tableline
HJD$^a$  &      $V$         &      $V$         &      $V$         &      $V$         &      $V$         \\
899.7859 & $22.892\pm0.029$ & $23.779\pm0.065$ & $23.254\pm0.040$ & $22.894\pm0.029$ & $23.816\pm0.067$ \\
899.8156 & $22.728\pm0.021$ & $23.803\pm0.055$ & $23.351\pm0.037$ & $22.993\pm0.027$ & $23.715\pm0.051$ \\
899.8390 & $22.687\pm0.022$ & $23.886\pm0.063$ & $23.506\pm0.045$ & $23.050\pm0.030$ & $24.165\pm0.083$ \\
899.8623 & $22.606\pm0.016$ & $23.711\pm0.041$ & $23.530\pm0.035$ & $23.155\pm0.025$ & $23.804\pm0.045$ \\
899.8861 & $22.619\pm0.013$ & $23.758\pm0.034$ & $23.677\pm0.032$ & $23.209\pm0.021$ & $23.800\pm0.035$ \\
\end{tabular}
\end{center}
\end{table}
\clearpage
\begin{table}
\begin{center}
Table \thetable---Continued \\
\begin{tabular}{cccccc}
\\
\tableline
\tableline
         &           C2-V06 &           C2-V07 &           C2-V08 &           C2-V09 &           C2-V10 \\
\tableline
HJD$^a$  &      $V$         &      $V$         &      $V$         &      $V$         &      $V$         \\
899.7859 & $23.786\pm0.065$ & $23.970\pm0.077$ & $23.296\pm0.042$ & $23.691\pm0.061$ & $23.041\pm0.034$ \\
899.8156 & $23.757\pm0.053$ & $24.050\pm0.069$ & $23.339\pm0.036$ & $23.646\pm0.048$ & $23.117\pm0.030$ \\
899.8390 & $23.744\pm0.056$ & $23.938\pm0.067$ & $23.164\pm0.033$ & $23.756\pm0.056$ & $23.124\pm0.032$ \\
899.8623 & $23.804\pm0.045$ & $24.054\pm0.055$ & $23.264\pm0.027$ & $23.589\pm0.037$ & $23.164\pm0.025$ \\
899.8861 & $23.725\pm0.033$ & $23.872\pm0.037$ & $23.271\pm0.022$ & $23.668\pm0.031$ & $23.229\pm0.021$ \\
\end{tabular}
\end{center}
\end{table}
\clearpage
\begin{table}
\begin{center}
Table \thetable---Continued \\
\begin{tabular}{cccccc}
\\
\tableline
\tableline
         &           C2-V11 &           C2-V12 &           C2-V13 &           C2-V14 &           C2-V15 \\
\tableline
HJD$^a$  &      $V$         &      $V$         &      $V$         &      $V$         &      $V$         \\
899.7859 & $22.654\pm0.024$ & $24.023\pm0.081$ & $23.530\pm0.052$ & $24.052\pm0.083$ & $24.342\pm0.108$ \\
899.8156 & $22.713\pm0.021$ & $23.793\pm0.055$ & $23.346\pm0.037$ & $24.071\pm0.070$ & $24.311\pm0.087$ \\
899.8390 & $22.856\pm0.026$ & $23.859\pm0.061$ & $23.324\pm0.038$ & $24.160\pm0.082$ & $24.331\pm0.094$ \\
899.8623 & $22.880\pm0.020$ & $24.066\pm0.056$ & $23.280\pm0.028$ & $23.950\pm0.051$ & $23.909\pm0.049$ \\
899.8861 & $22.971\pm0.017$ & $24.011\pm0.042$ & $23.356\pm0.024$ & $24.016\pm0.043$ & $23.590\pm0.029$ \\
\end{tabular}
\end{center}
\end{table}
\clearpage
\begin{table}
\begin{center}
Table \thetable---Continued \\
\begin{tabular}{cccccc}
\\
\tableline
\tableline
         &           C2-V16 &           C2-V17 &           C2-V18 &           C2-V19 &           C2-V20 \\
\tableline
HJD$^a$  &      $V$         &      $V$         &      $V$         &      $V$         &      $V$         \\
899.7859 & $23.219\pm0.039$ & $22.701\pm0.025$ & $23.198\pm0.039$ & $23.545\pm0.053$ & $23.939\pm0.077$ \\
899.8156 & $23.119\pm0.030$ & $22.644\pm0.020$ & $23.203\pm0.032$ & $23.469\pm0.041$ & $23.812\pm0.056$ \\
899.8390 & $23.130\pm0.032$ & $22.730\pm0.026$ & $23.304\pm0.038$ & $23.406\pm0.042$ & $23.901\pm0.065$ \\
899.8623 & $23.019\pm0.023$ & $22.729\pm0.017$ & $23.401\pm0.031$ & $23.435\pm0.032$ & $24.156\pm0.061$ \\
899.8861 & $22.972\pm0.017$ & $22.814\pm0.015$ & $23.424\pm0.026$ & $23.329\pm0.024$ & $23.933\pm0.040$ \\
\end{tabular}
\end{center}
\end{table}
\clearpage
\begin{table}
\begin{center}
Table \thetable---Continued \\
\begin{tabular}{cccccc}
\\
\tableline
\tableline
         &           C2-V21 &           C2-V22 &           C2-V23 &           C2-V24 &           C2-V25 \\
\tableline
HJD$^a$  &      $V$         &      $V$         &      $V$         &      $V$         &      $V$         \\
899.7859 & $23.489\pm0.050$ & $22.989\pm0.032$ & $23.579\pm0.056$ & $23.426\pm0.048$ & $23.150\pm0.038$ \\
899.8156 & $23.513\pm0.042$ & $22.913\pm0.025$ & $23.569\pm0.045$ & $23.467\pm0.041$ & $23.031\pm0.028$ \\
899.8390 & $23.541\pm0.046$ & $23.028\pm0.030$ & $23.557\pm0.048$ & $23.207\pm0.035$ & $22.863\pm0.026$ \\
899.8623 & $23.630\pm0.038$ & $22.932\pm0.021$ & $23.578\pm0.037$ & $22.989\pm0.022$ & $22.445\pm0.014$ \\
899.8861 & $23.649\pm0.031$ & $22.891\pm0.017$ & $23.541\pm0.029$ & $23.009\pm0.018$ & $22.247\pm0.010$ \\
\end{tabular}
\end{center}
\end{table}
\clearpage
\begin{table}
\begin{center}
Table \thetable---Continued \\
\begin{tabular}{cccccc}
\\
\tableline
\tableline
         &           C2-V26 &           C2-V27 &           C2-V28 &           C2-V29 &           C2-V30 \\
\tableline
HJD$^a$  &      $V$         &      $V$         &      $V$         &      $V$         &      $V$         \\
899.7859 & $23.917\pm0.074$ & $23.900\pm0.073$ & $23.274\pm0.042$ & $23.336\pm0.043$ & $23.170\pm0.038$ \\
899.8156 & $23.780\pm0.054$ & $23.857\pm0.057$ & $23.184\pm0.032$ & $23.096\pm0.029$ & $23.352\pm0.037$ \\
899.8390 & $23.849\pm0.062$ & $23.947\pm0.067$ & $23.238\pm0.035$ & $22.946\pm0.027$ & $23.394\pm0.042$ \\
899.8623 & $23.880\pm0.048$ & $24.076\pm0.057$ & $23.348\pm0.030$ & $22.983\pm0.022$ & $23.321\pm0.029$ \\
899.8861 & $23.864\pm0.038$ & $23.973\pm0.041$ & $23.374\pm0.024$ & $23.040\pm0.018$ & $23.442\pm0.026$ \\
\end{tabular}
\end{center}
\end{table}
\clearpage
\begin{table}
\begin{center}
Table \thetable---Continued \\
\begin{tabular}{cccccc}
\\
\tableline
\tableline
         &           C2-V31 &           C2-V32 &           C2-V33 &           C2-V34 &           C2-V35 \\
\tableline
HJD$^a$  &      $V$         &      $V$         &      $V$         &      $V$         &      $V$         \\
899.7859 & $22.963\pm0.032$ & $23.397\pm0.047$ & $25.114\pm0.218$ & $23.211\pm0.039$ & $23.610\pm0.056$ \\
899.8156 & $22.905\pm0.025$ & $23.283\pm0.035$ &      . . .       & $23.144\pm0.031$ & $23.625\pm0.047$ \\
899.8390 & $22.906\pm0.027$ & $23.238\pm0.036$ & $24.996\pm0.173$ & $23.044\pm0.030$ & $23.827\pm0.061$ \\
899.8623 & $22.958\pm0.021$ & $23.213\pm0.027$ & $25.004\pm0.132$ & $23.085\pm0.024$ & $23.782\pm0.044$ \\
899.8861 & $22.974\pm0.017$ & $23.031\pm0.018$ & $25.297\pm0.134$ & $23.126\pm0.020$ & $23.828\pm0.036$ \\
\end{tabular}
\end{center}
\end{table}
\clearpage
\begin{table}
\begin{center}
Table \thetable---Continued \\
\begin{tabular}{cccccc}
\\
\tableline
\tableline
         &           C2-V36 &           C2-V37 &           C2-V38 &           C2-V39 &           C2-V40 \\
\tableline
HJD$^a$  &      $V$         &      $V$         &      $V$         &      $V$         &      $V$         \\
899.7859 & $23.901\pm0.073$ & $23.480\pm0.050$ & $22.856\pm0.028$ & $22.551\pm0.022$ & $23.825\pm0.068$ \\
899.8156 & $23.823\pm0.056$ & $23.465\pm0.041$ & $22.687\pm0.021$ & $22.535\pm0.018$ & $23.461\pm0.041$ \\
899.8390 & $23.765\pm0.057$ & $23.347\pm0.040$ & $22.618\pm0.021$ & $22.573\pm0.020$ & $23.084\pm0.031$ \\
899.8623 & $23.832\pm0.046$ & $23.427\pm0.032$ & $22.538\pm0.015$ & $22.601\pm0.016$ & $22.884\pm0.020$ \\
899.8861 & $23.872\pm0.038$ & $23.551\pm0.029$ & $22.551\pm0.012$ & $22.667\pm0.013$ & $22.885\pm0.016$ \\
\end{tabular}
\end{center}
\end{table}
\clearpage
\begin{table}
\begin{center}
Table \thetable---Continued \\
\begin{tabular}{cccccc}
\\
\tableline
\tableline
         &           C2-V41 &           C2-V42 &           C2-V43 &           C2-V44 &           C2-V45 \\
\tableline
HJD$^a$  &      $V$         &      $V$         &      $V$         &      $V$         &      $V$         \\
899.7859 & $23.669\pm0.059$ & $22.454\pm0.020$ & $23.108\pm0.036$ & $23.491\pm0.050$ & $23.173\pm0.038$ \\
899.8156 & $23.749\pm0.053$ & $22.417\pm0.016$ & $23.024\pm0.028$ & $23.462\pm0.041$ & $23.032\pm0.028$ \\
899.8390 & $23.576\pm0.049$ & $22.388\pm0.017$ & $23.057\pm0.030$ & $23.524\pm0.045$ & $22.977\pm0.029$ \\
899.8623 & $23.642\pm0.038$ & $22.450\pm0.014$ & $22.857\pm0.020$ & $23.517\pm0.035$ & $23.022\pm0.023$ \\
899.8861 & $23.625\pm0.030$ & $22.417\pm0.011$ & $22.718\pm0.014$ & $23.535\pm0.028$ & $23.031\pm0.018$ \\
\end{tabular}
\end{center}
\end{table}
\clearpage
\begin{table}
\begin{center}
Table \thetable---Continued \\
\begin{tabular}{cccccc}
\\
\tableline
\tableline
         &           C2-V46 &           C2-V47 &           C2-V48 &           C2-V49 &           C2-V50 \\
\tableline
HJD$^a$  &      $V$         &      $V$         &      $V$         &      $V$         &      $V$         \\
899.7859 & $24.203\pm0.098$ & $23.189\pm0.038$ & $23.750\pm0.064$ & $24.409\pm0.115$ & $23.455\pm0.049$ \\
899.8156 & $23.767\pm0.054$ & $23.135\pm0.030$ & $23.789\pm0.055$ & $23.693\pm0.068$ & $23.480\pm0.041$ \\
899.8390 & $23.964\pm0.068$ & $23.102\pm0.032$ & $23.819\pm0.060$ & $23.989\pm0.070$ & $23.631\pm0.051$ \\
899.8623 & $23.815\pm0.046$ & $23.168\pm0.025$ & $23.792\pm0.045$ & $23.983\pm0.052$ & $23.597\pm0.037$ \\
899.8861 & $23.768\pm0.035$ & $23.197\pm0.021$ & $23.777\pm0.035$ & $23.992\pm0.042$ & $23.657\pm0.031$ \\
\end{tabular}
\end{center}
\end{table}
\clearpage
\begin{table}
\begin{center}
Table \thetable---Continued \\
\begin{tabular}{cccccc}
\\
\tableline
\tableline
         &           C2-V51 &           C2-V52 &           C2-V53 &           C2-V54 &           C2-V55 \\
\tableline
HJD$^a$  &      $V$         &      $V$         &      $V$         &      $V$         &      $V$         \\
899.7859 & $23.746\pm0.063$ & $23.859\pm0.071$ & $23.589\pm0.055$ & $23.796\pm0.066$ & $23.048\pm0.034$ \\
899.8156 & $23.595\pm0.046$ & $23.806\pm0.055$ & $23.677\pm0.050$ & $23.875\pm0.059$ & $23.040\pm0.028$ \\
899.8390 & $23.647\pm0.051$ & $23.863\pm0.062$ & $23.673\pm0.052$ & $23.782\pm0.058$ & $23.135\pm0.032$ \\
899.8623 & $23.725\pm0.041$ & $23.866\pm0.047$ & $23.789\pm0.044$ & $23.598\pm0.037$ & $23.253\pm0.027$ \\
899.8861 & $23.770\pm0.034$ & $23.900\pm0.038$ & $23.752\pm0.034$ & $23.499\pm0.027$ & $23.347\pm0.024$ \\
\end{tabular}
\end{center}
\end{table}
\clearpage
\begin{table}
\begin{center}
Table \thetable---Continued \\
\begin{tabular}{cccccc}
\\
\tableline
\tableline
         &           C2-V56 &           C2-V57 &           C2-V58 &           C2-V59 &           C2-V60 \\
\tableline
HJD$^a$  &      $V$         &      $V$         &      $V$         &      $V$         &      $V$         \\
899.7859 & $24.711\pm0.151$ & $23.575\pm0.054$ & $23.042\pm0.033$ &      . . .       & $22.658\pm0.024$ \\
899.8156 & $25.192\pm0.193$ & $23.354\pm0.037$ & $23.173\pm0.032$ & $23.854\pm0.058$ & $22.735\pm0.021$ \\
899.8390 & $25.163\pm0.202$ & $23.319\pm0.038$ & $23.085\pm0.031$ & $23.658\pm0.051$ & $22.783\pm0.024$ \\
899.8623 & $25.125\pm0.146$ & $23.232\pm0.027$ & $23.137\pm0.025$ & $23.616\pm0.038$ & $22.818\pm0.019$ \\
899.8861 & $25.243\pm0.129$ & $23.278\pm0.022$ & $23.080\pm0.019$ & $23.645\pm0.031$ & $22.872\pm0.016$ \\
\end{tabular}
\end{center}
\end{table}
\clearpage
\begin{table}
\begin{center}
Table \thetable---Continued \\
\begin{tabular}{cccccc}
\\
\tableline
\tableline
         &           C2-V61 &           C2-V62 &           C2-V63 &           C2-V64 &           C2-V65 \\
\tableline
HJD$^a$  &      $V$         &      $V$         &      $V$         &      $V$         &      $V$         \\
899.7859 & $23.475\pm0.050$ & $23.999\pm0.080$ & $22.761\pm0.026$ &      . . .       & $23.782\pm0.065$ \\
899.8156 & $23.487\pm0.042$ &      . . .       & $22.761\pm0.022$ & $22.764\pm0.022$ & $23.708\pm0.051$ \\
899.8390 & $23.306\pm0.038$ &      . . .       & $22.721\pm0.023$ & $22.825\pm0.025$ & $23.713\pm0.054$ \\
899.8623 & $23.333\pm0.029$ &      . . .       & $22.747\pm0.018$ & $22.759\pm0.018$ & $23.859\pm0.047$ \\
899.8861 & $23.265\pm0.022$ & $23.130\pm0.020$ & $22.707\pm0.014$ & $22.832\pm0.015$ & $23.776\pm0.035$ \\
\end{tabular}
\end{center}
\end{table}
\clearpage
\begin{table}
\begin{center}
Table \thetable---Continued \\
\begin{tabular}{cccccc}
\\
\tableline
\tableline
         &           C2-V66 &           C2-V67 &           C2-V68 &           C2-V69 &           C2-V70 \\
\tableline
HJD$^a$  &      $V$         &      $V$         &      $V$         &      $V$         &      $V$         \\
899.7859 & $25.165\pm0.231$ & $22.072\pm0.015$ & $23.428\pm0.047$ & $23.560\pm0.053$ & $23.246\pm0.041$ \\
899.8156 & $24.630\pm0.116$ & $22.054\pm0.012$ & $23.363\pm0.037$ & $23.317\pm0.036$ & $23.252\pm0.034$ \\
899.8390 & $24.988\pm0.171$ & $22.073\pm0.013$ & $23.530\pm0.046$ & $23.207\pm0.035$ & $23.252\pm0.036$ \\
899.8623 & $25.213\pm0.160$ & $22.088\pm0.010$ & $23.444\pm0.032$ & $23.219\pm0.026$ & $23.243\pm0.027$ \\
899.8861 & $25.623\pm0.181$ & $22.091\pm0.009$ & $23.529\pm0.028$ & $23.277\pm0.022$ & $23.327\pm0.024$ \\
\end{tabular}
\end{center}
\end{table}
\clearpage
\begin{table}
\begin{center}
Table \thetable---Continued \\
\begin{tabular}{cccccc}
\\
\tableline
\tableline
         &           C2-V71 &           C2-V72 &           C2-V73 &           C2-V74 &           C2-V75 \\
\tableline
HJD$^a$  &      $V$         &      $V$         &      $V$         &      $V$         &      $V$         \\
899.7859 & $23.305\pm0.042$ & $23.188\pm0.038$ & $23.531\pm0.052$ & $23.840\pm0.069$ & $24.327\pm0.109$ \\
899.8156 & $23.208\pm0.032$ & $23.310\pm0.036$ & $23.590\pm0.045$ & $23.681\pm0.049$ & $24.267\pm0.084$ \\
899.8390 & $23.218\pm0.034$ & $23.289\pm0.037$ & $23.580\pm0.048$ & $23.679\pm0.052$ & $24.102\pm0.077$ \\
899.8623 & $23.136\pm0.025$ & $23.442\pm0.032$ & $23.510\pm0.034$ & $23.715\pm0.041$ & $24.186\pm0.062$ \\
899.8861 & $23.081\pm0.019$ & $23.513\pm0.027$ &      . . .       & $23.827\pm0.036$ & $24.087\pm0.045$ \\
\end{tabular}
\end{center}
\end{table}
\clearpage
\begin{table}
\begin{center}
Table \thetable---Continued \\
\begin{tabular}{cccccc}
\\
\tableline
\tableline
         &           C2-V76 &           C2-V77 &           C2-V78 &           C2-V79 &           C2-V80 \\
\tableline
HJD$^a$  &      $V$         &      $V$         &      $V$         &      $V$         &      $V$         \\
899.7859 & $23.355\pm0.044$ & $23.157\pm0.038$ & $23.729\pm0.062$ &      . . .       &      . . .       \\
899.8156 & $23.258\pm0.034$ & $23.330\pm0.036$ & $23.221\pm0.033$ & $24.756\pm0.130$ & $25.698\pm0.309$ \\
899.8390 & $23.117\pm0.032$ & $23.404\pm0.041$ & $22.820\pm0.025$ &      . . .       &      . . .       \\
899.8623 & $23.156\pm0.025$ & $23.464\pm0.033$ & $22.822\pm0.019$ & $24.819\pm0.110$ &      . . .       \\
899.8861 & $23.168\pm0.020$ & $23.580\pm0.029$ & $22.874\pm0.016$ & $24.828\pm0.088$ & $25.445\pm0.153$ \\
\end{tabular}
\end{center}
\end{table}
\clearpage
\begin{table}
\begin{center}
Table \thetable---Continued \\
\begin{tabular}{cccccc}
\\
\tableline
\tableline
         &           C2-V81 &           C2-V82 &                  &                  &                  \\
\tableline
HJD$^a$  &      $V$         &      $V$         &                  &                  &                  \\
899.7859 & $24.201\pm0.095$ & $23.177\pm0.038$ &                  &                  &                  \\
899.8156 & $24.389\pm0.094$ & $23.179\pm0.031$ &                  &                  &                  \\
899.8390 & $24.232\pm0.087$ & $23.316\pm0.038$ &                  &                  &                  \\
899.8623 & $24.071\pm0.056$ & $23.323\pm0.029$ &                  &                  &                  \\
899.8861 & $23.746\pm0.033$ & $23.350\pm0.024$ &                  &                  &                  \\
\end{tabular}
\end{center}
\end{table}

\clearpage
\begin{deluxetable}{lrrrrrrrl}
\tablecaption{Variable Stars. \label{tab_variables}}
\tablewidth{0pt}
\tablehead{
\colhead{ID\tablenotemark{a}} &
\colhead{Chip} &
\colhead{X} &
\colhead{Y} &
\colhead{V} &
\colhead{R} &
\colhead{P (d)} &
\colhead{Q\tablenotemark{b}} &
\colhead{Class\tablenotemark{c}}}
\startdata
C1-V01 & 1 &  691.71 &  844.00 & $25.38 \pm 0.11$ & $25.21 \pm 0.23$ & $0.11 \pm 0.01$ & 2 & RR \\
C1-V02 & 1 & 1443.33 & 2494.99 & $23.75 \pm 0.06$ & $23.66 \pm 0.10$ & $0.59 \pm 0.02$ & 3 & C-FO \\
C1-V03 & 1 & 1561.26 & 2010.57 & $25.03 \pm 0.09$ & $24.78 \pm 0.06$ & $0.61 \pm 0.02$ & 3 & RRab \\
C1-V04 & 1 & 1663.61 & 2872.98 & $24.02 \pm 0.07$ & $23.76 \pm 0.03$ & $0.44 \pm 0.01$ & 2 & C \\
C1-V04 & 1 & 1663.61 & 2872.98 & $24.18 \pm 0.11$ & $23.76 \pm 0.03$ & $0.86 \pm 0.04$ & 2 & C \\
C1-V05 & 1 & 1712.41 & 3989.46 & $23.56 \pm 0.09$ & $23.17 \pm 0.12$ & $1.58 \pm 0.21$ & 4 & C-FM? \\
C1-V06 & 1 & 1735.66 & 1215.41 & $23.44 \pm 0.08$ & $23.27 \pm 0.15$ & $0.61 \pm 0.02$ & 4 & C-FO \\
C1-V07 & 1 & 1761.25 & 2635.05 & $25.23 \pm 0.13$ & $25.01 \pm 0.03$ & $0.43 \pm 0.01$ & 3 & RRc \\
\enddata
\tablenotetext{a}{Stars listed multiple times have periods with multiple potential aliases.}
\tablenotetext{b}{Variable star quality is based on the cleanness of the light curves, uniqueness of the period, and coherence between $V$ and $R$ light curves, where 4 is highest quality and 0 is lowest quality.}
\tablenotetext{c}{RR Lyraes classified into subclasses (RRab, RRc, RRd) when possible.  Cepheid classifications (FM = fundamental mode, FO = first overtone, SO = second overtone) made only for objects with unambiguous periods and quality ratings of 3 or 4.}
\tablenotetext{d}{Upper limits of the periods of stars C2-V08, C2-V63, and C2-V64 are unconstrained.}
\end{deluxetable}
%paste ../phot/variables/combine/final2.txt ../phot/variables/mags.output_proc | awk 'NR>1 {printf "%s & 2 & %7.2f & %7.2f & $%5.2f \\pm %4.2f$ & $%5.2f \\pm %4.2f$ & $%4.2f \\pm %4.2f$ \\\\\n",$1,$8,$9,$23,$24,$28,$29,$17,$18}'

\end{document}